\documentclass[journal]{IEEEtran}
\usepackage{mathrsfs}
\usepackage{bbm}
\usepackage{amssymb}
\usepackage{bbding}
\usepackage{threeparttable}
\usepackage{colortbl}
\usepackage[mathcal]{euscript}
\usepackage{psfrag,calc,url,bm}
\usepackage{cite}
\usepackage{graphicx}
\usepackage{psfrag}
\usepackage{subfigure}
\usepackage{hyperref}
\usepackage{url}
\usepackage{stfloats}
\usepackage{amsmath}
\usepackage{float}
\usepackage{algorithmic}
\usepackage[ruled,linesnumbered,vlined]{algorithm2e}
\usepackage{color}
\usepackage{boxedminipage}
\usepackage{amsthm}
\usepackage{multirow}
\usepackage{setspace}
\usepackage{soul}
\usepackage{epstopdf}
\usepackage{svg}
\allowdisplaybreaks[3]
\usepackage{multirow}
\usepackage{amsmath}
\usepackage{booktabs}

\usepackage{makecell}
\usepackage{diagbox}

\begin{document}

\title{Graph Neural Network Enabled Fluid Antenna Systems: A Two-Stage Approach}
\author{Changpeng He, Yang Lu,~\IEEEmembership{Member,~IEEE}, Wei Chen~\IEEEmembership{Senior Member,~IEEE}, Bo Ai,~\IEEEmembership{Fellow,~IEEE}, \\Kai-Kit Wong,~\IEEEmembership{Fellow,~IEEE}, and Dusit Niyato,~\IEEEmembership{Fellow,~IEEE}
\thanks{Changpeng He and Yang Lu are with the State Key Laboratory of Advanced Rail Autonomous Operation, and also with the School of Computer Science and Technology, Beijing Jiaotong University, Beijing 100044, China (e-mail: 24120332@bjtu.edu.cn, yanglu@bjtu.edu.cn).}
\thanks{Wei Chen and Bo Ai are with the School of Electronics and Information Engineering, Beijing Jiaotong University, Beijing 100044, China (e-mail: weich@bjtu.edu.cn, boai@bjtu.edu.cn).}
\thanks{Kai-Kit Wong is with the Department of Electronic and Electrical Engineering, University College London, WC1E 6BT London, U.K., and also with the Yonsei Frontier Laboratory and the School of Integrated Technology, Yonsei  University, Seoul 03722, South Korea (e-mail: kat-kit.wong@ucl.ac.uk).}
\thanks{Dusit Niyato is with the College of Computing and Data Science, Nanyang Technological University, Singapore 639798 (e-mail:  dniyato@ntu.edu.sg).}
}
\maketitle

\begin{abstract}
An emerging fluid antenna system (FAS) brings a new dimension, i.e., the antenna positions, to deal with the deep fading, but simultaneously introduces challenges related to the transmit design. This paper proposes an ``unsupervised learning to optimize" paradigm to optimize the FAS. Particularly, we formulate the sum-rate and energy efficiency (EE) maximization problems for a multiple-user multiple-input single-output (MU-MISO) FAS and solved by a two-stage graph neural network (GNN) where the first stage and the second stage are for the inference of antenna positions and beamforming vectors, respectively. The outputs of the two stages are jointly input into a unsupervised loss function to train the two-stage GNN. The numerical results demonstrates that the advantages of the FAS for performance improvement and the two-stage GNN for real-time and scalable optimization. Besides, the two stages can function separately.
\end{abstract}
\begin{IEEEkeywords} FAS, unsupervised learning, MU-MISO, two-stage GNN.
\end{IEEEkeywords}

\section{Introduction}

Channel conditions are the key factor for wireless signal propagation. The spatial degree of freedom introduced by  multiple-input  multiple-output  enhances the channel conditions with diversity and multiplexing gains. Nevertheless, traditional wireless systems regard the channel conditions uncontrollable and thus, mainly focus on the optimization of transceivers such as transmit and receive precoding. Recently, the reconfigurable antenna technologies have been proposed to realize programmable channel conditions, which provides a new dimension for wireless optimization. Particularly, the fluid antenna system (FAS) \cite{fas} motivated by liquid metals switches the antennas over a fixed-length line space to deal with the deep fading. Therefore, the FAS allows jointly optimization of transceivers and channel conditions, whose capability to boost the capacity has been validated in various scenarios \cite{fas2}.

On the other hand,  jointly optimization of transceivers and channel conditions  struggles traditional convex optimization techniques. The deep learning (DL) has been emerged as a new optimization paradigm to address complicated problems \cite{aifas}, for example, the multilayer perceptron (MLP) was adopted to jointly optimize the antenna positions and multicast beamforming for the FAS \cite{FAS-MLP}. However, the MLP may be with limited generalization performance. Recently, the graph neural network (GNN) has drawn increasing attention for wireless networks with graph-structured topology \cite{lugnn}. In \cite{dl,dl2,dl3}, the GNNs based on the multi-head attention mechanism and the residual connection were proposed to address the energy efficiency (EE) maximization, sum-rate maximization and max-min rate problems for multiple-user multiple-input single-output (MU-MIMO) systems, respectively. In \cite{twostagegnn}, the placement and transmission design of the unmanned aerial vehicle were jointly optimized by a two-stage GNN. In \cite{icnet} and \cite{new2}, the GNNs were applied to interference channels and physical-layer scenarios. The aforementioned works  showed that the GNN outperformed the MLP and achieved near-optimal and real-time inference.

To the best of our knowledge, the GNN-enabled FAS design has not been reported thus far. To fill this gap, this paper formulate a joint antenna placement and beamforming optimization problem in terms of sum-rate and EE maximization. We propose a two-stage GNN with the first stage and the second stage to obtain antenna positions and beamforming vectors, respectively. In both stages, the multi-head attention mechanism and residual connection are adopted, and in the second stage, we utilize the hybrid maximum ratio transmission (MRT) and zero forcing (ZF) scheme \cite{hyb}. The dedicated activation functions are design to guarantee feasible solution, and the two-stage GNN is trained unsupervisedly. Numerical results demonstrate advantages of the FAS for performance improvement and the effectiveness of the proposed two-stage GNN in terms of optimality, scalability and computational efficiency. Besides, the two stages jointly trained also function separately, and can be integrated with traditional algorithms.

\section{System Model and Problem Definition}

\begin{figure}[t]
{\centering
{\includegraphics[ width=.44\textwidth]{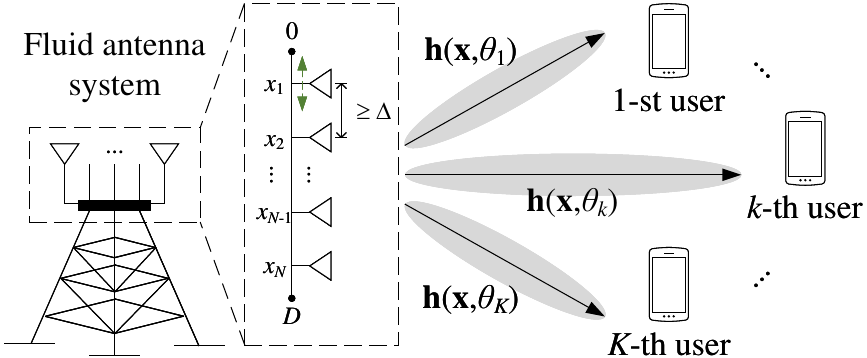}}}
\caption{Illustration of MU-MISO FAS.}
\label{sys}
\end{figure}

As shown in Fig. \ref{sys}, we consider a MU-MISO system where a base station (BS) equipped with a FAS serves $K$ users. The FAS at the BS includes $N$ fluid antennas that are linearly distributed while each user is equipped with single fixed-position antenna. The positions of the $N$ fluid antennas at the BS are adjustable along a line segment of length $D$. Let $x_n$ $(n \in {\cal N}\triangleq\{1, \ldots, N\})$ represent the position of the $n$-th fluid antenna. Without loss of generality, we consider that $0 \leq x_1 < x_2 < \ldots < x_{N} \leq D$. Denote the steering angle from the FAS to the $k$-th $(k \in {\cal K}\triangleq\{1, \ldots, K\})$ user as $\theta_k \in [0,\pi)$, and the corresponding steer vector is given by
\begin{flalign}\label{channel}
&{\bf h}({\bf x}, \theta_k) = \nonumber\\
&\left[ e^{{\rm i} \frac{2\pi}{\lambda} x_1 \cos\left(\theta_k\right)}, e^{{\rm i} \frac{2\pi}{\lambda} x_2 \cos\left(\theta_k\right)}, \dots, e^{{\rm i} \frac{2\pi}{\lambda} x_{N} \cos\left(\theta_k\right)} \right]^T,
\end{flalign}
where ${\bf x} \triangleq [x_1, x_2, \ldots, x_{N}]^{{T}}$ represents the antenna position vector and $\lambda$ denotes the wavelength.

\begin{figure}[t]
{\centering
{\includegraphics[ width=.48\textwidth]{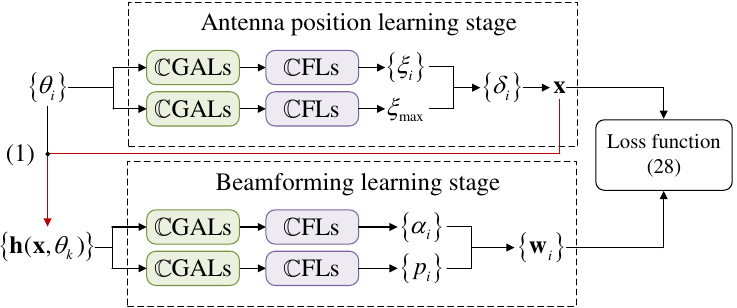}}}
\caption{The structure of the two-stage GNN.}
\label{structure}
\end{figure}

The BS transmits $K$ different information symbols for the $K$ users over the same spectral bandwidth. Denote the beamforming vector for the $k$-th user as ${\bf w}_k \in {\mathbb C}^N$, and the signal-to-interference-plus-noise ratio (SINR) observed at the $k$-th user is given by
\begin{flalign}
& \gamma_k \left(\left\{\mathbf{w}_i,\mathbf{x}\right\}\right) = \frac{\left|{\bf w}_k^H \sqrt{d_k}{\bf h}\left({\bf x}, \theta_k\right)\right|^2}{\sum_{i\in {\cal K}\setminus k}\left|{\bf w}_i^H \sqrt{d_k}{\bf h}\left({\bf x}, \theta_k\right)\right|^2 + \sigma_k^2},
\end{flalign}
where $d_k$ and $\sigma_k^2$ denote the large-scale path loss and the noise power of the $k$-th user, respectively.

Our goal is to maximize the system utility under constraints of the transmit power budget and the adjustable area of the fluid antennas. Then, the considered problem is formulated as
\begin{subequations}\label{po2}
\begin{align}
&\max_{\left\{\mathbf{w}_i,\mathbf{x}\right\}}~U\left(\left\{\mathbf{w}_i,\mathbf{x}\right\}\right)\\
\text{s.t.}~& \sum\nolimits_{i\in {\cal K}} \left\|\mathbf{w}_i\right\|^2 \leq P_{\max},\label{p1:2b} \\
& x_1 \geq 0,~x_{N} \leq D,\label{p1:2c}\\
&x_n - x_{n-1} \geq \Delta,~n\in{\cal N}\setminus\left\{1\right\}, \label{p1:2d}
\end{align}
\end{subequations}
where $P_{\max}$ denotes the power budget, $\Delta > 0$ denotes the minimum spacing between any two adjacent fluid antennas to avoid coupling effects and the system utility function is the sum rate, i.e.,
\begin{flalign}\label{suf1}
   U\left(\left\{\mathbf{w}_i,\mathbf{x}\right\}\right) =\sum\nolimits_{k \in {\cal K}} \log_2\left(1+\gamma_k \left(\left\{\mathbf{w}_i,\mathbf{x}\right\}\right)\right),
\end{flalign}
or EE, i.e,
\begin{flalign}\label{suf2}
   U\left(\left\{\mathbf{w}_i,\mathbf{x}\right\}\right) =\frac{\sum\nolimits_{k \in {\cal K}}~\log_2\left(1+\gamma_k \left(\left\{\mathbf{w}_i,\mathbf{x}\right\}\right)\right)}{\sum\nolimits_{i\in {\cal K}} \left\|\mathbf{w}_i\right\|^2 +P_{\rm c}},
\end{flalign}
where $P_{\rm c}$ denotes the constant power consumption at the BS.

Problem \eqref{po2} is challenging to address by the traditional convex techniques due to the deeply coupled $\{{\bf w}_i\}$ and $\bf x$ in $\gamma_k (\{\mathbf{w}_i,\mathbf{x}\})$. Instead, this paper intend to solve Problem \eqref{po2} following the ``unsupervised learning to optimize" paradigm.

\section{Two-stage GNN}

In order to leverage the graph-structured topology  of wireless networks while handling Problem \eqref{po2}, we convert the considered system into a fully connected directed graph denoted by $\mathcal{G} = (\mathcal{V},\mathcal{E})$. Here,  $\mathcal{V}$ with $|\mathcal{V}|=K$ denotes the set of nodes, each of which represents one user, and $\mathcal{E}$ denotes the set of edges, each of which represents the relationship (including interference, sharing power budget, etc.) between two nodes. Our goal is to construct and train a DL model to map the graph to a (near-)optimal solution to Problem \eqref{po2}.

As illustrated in Fig. \ref{structure}, the proposed DL model includes two stages, i.e., the antenna position learning stage and the beamforming learning stage. Both two stages are based on GNN and include two types of layers, i.e., the complex graph attention layer (${\mathbb C}$GAL) and the complex fully-connected layer (${\mathbb C}$FL). In the follows, we first gives the structure of the two-stage GNN, and then, introduces the two layers.

\subsection{The Structure of the Two-stage GNN}

\subsubsection{Antenna position learning stage}

The input and output of the antenna position learning stage are $\{{\theta_i}\}$ and $\bf{x}$, respectively. The $k$-th node is with feature of $\theta_k$.

To guarantee feasible $\bf{x}$, the following reformulation is adopted. We define auxiliary variables
\begin{flalign}
&{\delta _1} = 0,~{\delta _n}=x_n - x_{n-1}-\Delta,~n\in{\cal N}\setminus\left\{1\right\}, \label{delta}\\
&{\delta _{\max }} = D - \left( {N - 1} \right)\Delta,
\end{flalign}
and we re-express \eqref{p1:2c} and \eqref{p1:2d} as
\begin{flalign}\label{new_cons}
{x_1} = 0,~{\delta _n} \ge 0,~\sum\nolimits_{n \in {\cal N} } {{\delta _n}}  \le {\delta _{\max }}.
\end{flalign}
Following \cite{FAS-MLP}, there is no optimality loss by setting ${x_1} = 0$. Then, define auxiliary variables $\{\xi _i\}$ and ${\xi _{\max }}$, ${\delta _n}$ can be further represented by
\begin{flalign}\label{delta_n}
{\delta _n} = \frac{{{\xi _n}}}{{\sum\nolimits_{i\in{\cal N}\setminus\left\{1\right\}} {{\xi _i}} }} \times \frac{{{\delta _{\max }}}}{{1 + {e^{ - {\xi _{\max }}}}}},~n\in{\cal N}\setminus\left\{1\right\},
\end{flalign}
Note that regardless of the values of $\{\xi _i\}$ and ${\xi _{\max }}$, ${\delta _n}$ by \eqref{delta_n} satisfies \eqref{new_cons}. Therefore, we learn the unconstrained $\{\xi _i\}$ and ${\xi _{\max }}$ instead, and recover $\{\delta_i\}$ following \eqref{delta_n} to obtain ${\bf x}$ by
\begin{flalign}
{x_n} = (n - 1)\Delta  + \sum\nolimits_{i \in {\cal N}}^n {{\delta _i}}, \forall {n\in\cal N},
\end{flalign}
which is feasible to Problem \eqref{po2}.

Particularly, we use $G_1$ ${\mathbb C}$GALs and $F_1$ ${\mathbb C}$FLs to construct the mapping from from $\{{\theta_i}\}$ to $\{\xi _i|i\in{\cal N}\setminus\left\{1\right\}\}$, while using $G_2$ ${\mathbb C}$GALs and $F_2$ ${\mathbb C}$FLs to construct the mapping from $\{{\theta_i}\}$ to ${\xi _{\max }}$.

With the obtained ${\bf x}$, $\{{\bf h}({\bf x}, \theta_i)\}$ are calculated through \eqref{channel}, which are input into the beamforming learning stage via a residual connection (marked as red color in Fig. \ref{structure}).

\subsubsection{Beamforming learning stage}
The input and output of the beamforming learning stage are $\{{\bf h}\left({\bf x}, \theta_i\right)\}$ and $\{{\bf{w}}_i\}$ respectively. The $k$-th node is with feature of ${\bf h}({\bf x},{\theta_k})$.

We adopt a model-based learning approach, i.e., HZF learning \cite{hyb}, which separate the $k$-th beamforming vector as its power part denoted by $p_k$ and its  direction part denoted by ${{\overline{\bf{w}}}_k}$, i.e.,
\begin{flalign}\label{mmse}
{{\bf{w}}_k} = {\sqrt {p_k}}{{\overline{\bf{w}}}_k}\left(\alpha_k\right),~{\left\| {{{\overline {\bf{w}} }_k}} \left(\alpha_k\right)\right\|^2} = 1,
\end{flalign}
where
\begin{flalign}\label{hybrid}
{\overline {\bf w}}_k \left(\alpha_k\right) = \frac{\alpha_k \frac{{\bf u}_k}{\|{\bf u}_k\|} + \left(1-\alpha_k\right)\frac{{\bf h}\left({\bf x}, \theta_k\right)}{\|{\bf h}\left({\bf x}, \theta_k\right)\|}}{\left\|\alpha_k \frac{{\bf u}_k}{\|{\bf u}_k\|} + \left(1-\alpha_k\right)\frac{{\bf h}\left({\bf x}, \theta_k\right)}{\|{\bf h}\left({\bf x}, \theta_k\right)\|}\right\|},
\end{flalign}
where $\alpha_k \in [0,1]$ denotes the hybrid coefficient, and ${\bf u}_k$ is the $k$-th column of
\begin{flalign}\label{U_q}
{{\bf{U}}_k} = {\bf{G}}_k^H{\left( {{{\bf{G}}_k}{\bf{G}}_k^H} \right)^{ - 1}},
\end{flalign}
where ${\bf G}_k\triangleq[{\bf h}^H({\bf x}, \theta_1),{\bf h}^H({\bf x}, \theta_2),\ldots,{\bf h}^H({\bf x}, \theta_K)]$. Note that the HZF learning may slightly degrade the optimality, but it significantly simplifies the output from $\{{\bf w}_i\}$ to $\{{p}_i,\alpha_i\}$, which helps to enhance the expressive performance of the DL model handling Problem \eqref{po2}. Therefore, we learn $\{{p}_i,\alpha_i\}$ instead, and recover $\{{\bf w}_i\}$ following \eqref{hybrid}.

Particularly, we use $G_3$ ${\mathbb C}$GAL layers and $F_3$ ${\mathbb C}$FL to construct the mapping from $\{{\bf h}({\bf x}, \theta_i)\}$ to $\{\alpha_i\}$, while using $G_4$ ${\mathbb C}$GALs  and $F_4$ ${\mathbb C}$FLs to construct the mapping from $\{{\bf h}({\bf x}, \theta_i)\}$ to  $\{p_i\}$.

Besides, to guarantee feasible $\{{\bf w}_i\}$, we use the Sigmoid activation function to output $\alpha_k$ being in $(0,1)$, while using the following activation function to output
\begin{flalign}
{p_k} := \left\{ {\begin{array}{*{20}{l}}
{{p_k},~\sum\nolimits_{i \in {\mathcal K}} {{p_i}}  \le {P_{\max }}}\\
{\frac{{{p_k}}}{{\sum\nolimits_{i \in {\mathcal K}} {{p_i}} }}P_{\max },~\sum\nolimits_{i \in {\mathcal K}} {{p_i}}  > {P_{\max }}}
\end{array}} \right.,
\end{flalign}
such that $\sum_{i\in{\cal K}}\|{{{\bf{w}}}_i}\|^2=\sum_{i\in{\cal K}}p_i\le P_{\max}$.

\subsection{Complex Graph Attention Layer}

The ${\mathbb C}$GAL employs multi-head attention mechanism to facilitate each node extract graphical information via message passing. Besides, the residual structure via the skip connection is utilized to mitigate the over-smoothing issue. For the first stage, a virtual node is added to aggregate features from other nodes, since $\bf{x}$ is a graph-level output. For the second stage, the node-level output is adopted.

Consider that $G$ ($G\in\{G_1,G_2,G_3,G_4\}$) ${\mathbb C}$GALs are employed. Denote ${\bf V}^{({{\mathbb C}\rm
 GAL})}_{(g-1)}\in {\mathbb C}^{K\times V_{(g-1)}}$ by the output/input node feature matrix of the $(g-1)$-th/$g$-th ($g\in \{1,2,\ldots,G\}$) ${\mathbb C}$GAL, where $V_{g}$ represents the feature dimension of each node.  Particularly, the input feature of the $k$-th node of the $1$-st ${\mathbb C}$GAL  ${\bf V}^{({{\mathbb C}\rm
 GAL})}_{0}$ is given by
 \begin{flalign}
{\left[ {{\bf{V}}_0^{({{\mathbb C}\rm{GAL}})}} \right]_{k,:}} =\left\{ {\begin{array}{*{20}{l}}
 {\theta _k},~{\rm First~stage}\\
 {\bf{h}}\left( {{\bf{x}},{\theta _k}} \right),~{\rm Second~stage}
\end{array}} \right..
\end{flalign}

 Each ${\mathbb C}$GAL employs $Z$ attention heads. The coefficient of the $z$-th ($z\in\{1,2,\ldots,Z\}$) attention head in the $g$-th
${\mathbb C}$GAL is denoted by ${\bf A}^{({{\mathbb C}\rm
 GAL})}_{g,z}\in {\mathbb R}^{K\times K}$, which is given by (\ref{attention coefficient0}), where ${\rm LeakyReLU}(\cdot)$ represents the LeakyReLU activation function, ${\rm Re}(\cdot)$ represents the real part, and ${\bf W}^{({{\mathbb C}\rm
 GAL})}_{g,z}\in {\mathbb C}^{V_{(g-1)}\times V_{g}}$  and ${\bf a}^{({{\mathbb C}\rm
 GAL})}_{g,z}\in {\mathbb C}^{2V_{g}}$ denote the learnable parameters associated with the input node feature and attention head, respectively. Then, ${\bf V}^{({{\mathbb C}\rm GAL})}_{g}$ is calculated by \eqref{update}, where ${\rm {\mathbb C}ReLU}(\cdot)$ represents the complex ReLU activation function, and ${\widehat{\bf W}}^{({{\mathbb C}\rm GAL})}_g\in {\mathbb C}^{V_{(g-1)}\times V_{g}}$ denotes the learnable parameters associated with the residual connection.

\begin{figure*}
\begin{flalign}
\label{attention coefficient0}
&{[{\bf{A}}_{g,z}^{({{\mathbb C}\rm GAL})}]_{k,j}} = \frac{{\exp \left( {{{\rm{LeakyReLU}}}\left( {{\mathop{\rm Re}\nolimits} \left( {{\bf{a}}{{_{g,z}^{({{\mathbb C}\rm GAL})}}^T}\left( {{{\left[ {{\bf{V}}_{(g - 1)}^{({{\mathbb C}\rm GAL})}} \right]}_{k,:}}{\bf{W}}_{g,z}^{({{\mathbb C}\rm GAL})}||{{\left[ {{\bf{V}}_{(g - 1)}^{({{\mathbb C}\rm GAL})}} \right]}_{j,:}}{\bf{W}}_{g,z}^{({{\mathbb C}\rm GAL})}} \right)} \right)} \right)} \right)}}{{\sum\nolimits_{i \in {\mathcal V}} {\exp } \left( {{{\rm{LeakyReLU}}}\left( {{\mathop{\rm Re}\nolimits} \left( {{\bf{a}}{{_{g,z}^{({{\mathbb C}\rm GAL})}}^T}\left( {{{\left[ {{\bf{V}}_{(g - 1)}^{({{\mathbb C}\rm GAL})}} \right]}_{k,:}}{\bf{W}}_{g,z}^{({{\mathbb C}\rm GAL})}||{{\left[ {{\bf{V}}_{(g - 1)}^{({{\mathbb C}\rm GAL})}} \right]}_{i,:}}{\bf{W}}_{g,z}^{({{\mathbb C}\rm GAL})}} \right)} \right)} \right)} \right)}}
\end{flalign}
\hrule
\end{figure*}

\begin{figure*}
\begin{flalign}\label{update}
{\left[ {{\bf{V}}_g^{({{\mathbb C}\rm
 GAL})}} \right]_{k,:}}=& \mathop {||}\limits_{z = 1}^{Z}  {\rm {\mathbb C}ReLU}\left(\sum\nolimits_{j \in {\mathcal V}} [{\bf A}^{({{\mathbb C}\rm
 GAL})}_{g,z}]_{k,j} {\left[ {{\bf{V}}_{g-1}^{({{\mathbb C}\rm
 GAL})}} \right]_{j,:}}{\bf W}^{({{\mathbb C}\rm
 GAL})}_{g,z} + {\left[ {{\bf{V}}_{(g-1)}^{({{\mathbb C}\rm
 GAL})}} \right]_{k,:}} {\widehat {\bf W}}^{({{\mathbb C}\rm
 GAL})}_{g,z} \right)
\end{flalign}
\hrule
\end{figure*}

As mentioned, in the first stage, there is a virtual node connecting other nodes. Therefore, each ${\mathbb C}$GAL layer employs \eqref{attention coefficient2} to obtain the attention coefficient ${{\bm{\beta}}_{g,z}^{({{\mathbb C}\rm{GAL}})}}\in {\mathbb R}^{K}$ for the virtual node, where ${\overline {\bf W}}^{({{\mathbb C}\rm
 GAL})}_{g,z}\in {\mathbb C}^{V_{g}\times V_{g}}$  and ${\overline {\bf a}}^{({{\mathbb C}\rm
 GAL})}_{g,z}\in {\mathbb C}^{V_{g}}$ denote the  learnable parameters.  Then, the feature of the virtual node denoted by ${\bf{v}}_g^{({{\mathbb C}\rm{GAL}})}\in {\mathbb C}^{V_{g}}$ is updated by \eqref{vn_update}, where  ${\bf {\widetilde W}}^{({{\mathbb C}\rm
 GAL})}_{g,z}\in {\mathbb C}^{V_{(g-1)}\times V_{g}}$ denotes the learnable parameters for residual connection.

\begin{figure*}
\begin{flalign}
\label{attention coefficient2}
&{[{\bm{\beta }}_{g,z}^{({{\mathbb C}\rm{GAL}})}]_j} = \frac{{\exp \left( {{{\rm{LeakyReLU}}}\left( {{\mathop{\rm Re}\nolimits} \left( {{\bf{\overline a}}{{_{g,z}^{({{\mathbb C}\rm{GAL}})}}^T}{{\left[ {{\bf{V}}_g^{({{\mathbb C}\rm{GAL}})}} \right]}_{j,:}}{\bf{\overline W}}_{g,z}^{({{\mathbb C}\rm{GAL}})}} \right)} \right)} \right)}}{{\sum\nolimits_{i \in {\mathcal V}} {\exp } \left( {{{\rm{LeakyReLU}}}\left( {{\mathop{\rm Re}\nolimits} \left( {{\bf{\overline a}}{{_{g,z}^{({{\mathbb C}\rm{GAL}})}}^T}{{\left[ {{\bf{V}}_g^{({{\mathbb C}\rm{GAL}})}} \right]}_{i,:}}{\bf{\overline W}}_{g,z}^{({{\mathbb C}\rm{GAL}})}} \right)} \right)} \right)}}
\end{flalign}
\hrule
\end{figure*}

\begin{figure*}
\begin{flalign}\label{vn_update}
{\bf{v}}_g^{({{\mathbb C}\rm{GAL}})} = \mathop {||}\limits_{z = 1}^Z {\rm{ReLU}}\left( {\sum\nolimits_{j \in {\mathcal V}} {{{[{\bm{\beta }}_{g,z}^{({{\mathbb C}\rm{GAL}})}]}_j}{{\left[ {{\bf{V}}_g^{({{\mathbb C}\rm{GAL}})}} \right]}_{j,:}}{\bf{\overline W}}_{g,z}^{({{\mathbb C}\rm{GAL}})} + {\bf{ v}}_{\left( {g - 1} \right)}^{({{\mathbb C}\rm{GAL}})}{\bf{\widetilde W}}_{g,z}^{({{\mathbb C}\rm{GAL}})}} } \right)
\end{flalign}
\hrule
\end{figure*}

\subsection{Complex Fully-connected Layer}

The ${\mathbb C}$FL decodes the features extracted  by the ${\mathbb C}$GALs to the desired solution, i.e., $\{\xi _i\}$, ${\xi _{\max }}$, $\{\alpha_i\}$ and $\{p_i\}$.

Consider that $F$ ($G\in\{G_1,G_2,G_3,G_4\}$) ${\mathbb C}$FLs are employed. Denote ${\bf V}^{({{\mathbb C}\rm FL})}_{(f-1)}$ (which can be a matrix of ${\mathbb C}^{K\times H_{(f-1)}}$ or a vector of
${\mathbb C}^{H_{(f-1)}}$) by the output/input node feature matrix of the $(f-1)$-th/$f$-th ($f\in\{1,2,\ldots,F\}$) ${\mathbb C}$FL, where $H_{f}$ represents the feature dimension of each node. Particularly, the input node feature matrix of the $1$-st ${\mathbb C}$FL ${\bf V}^{({{\mathbb C}\rm FL})}_{0}$ is given by
 \begin{flalign}
{\bf{V}}_0^{({{\mathbb C}\rm{FL}})} =\left\{ {\begin{array}{*{20}{l}}
 {\bf{v}}_G^{({{\mathbb C}\rm{GAL}})},~{\rm First~stage}\\
 {\bf{V}}_G^{({{\mathbb C}\rm{GAL}})},~{\rm Second~stage}
\end{array}} \right..
\end{flalign}

The  $f$-th ${\mathbb C}$FL with $(f\in \{1,2,\ldots,F-1\})$ updates the node feature matrix by
\begin{flalign}\label{cfl}
{{\bf V}^{({{\mathbb C}\rm FL})}_{f} = {\mathbb C}{\rm ReLU} \left({\bf V}^{({{\mathbb C}\rm FL})}_{\left(f-1\right)} {\bf W}^{({{\mathbb C}\rm FL})}_f + {\bf B}^{({{\mathbb C}\rm FL})}_f\right)},
\end{flalign}
where ${\bf W}^{({{\mathbb C}\rm FL})}_f \in {\mathbb C}^{H_{(f-1)}\times H_{f}}$ and ${\bf B}^{({{\mathbb C}\rm FL})}_f$ (which can be a matrix of ${\mathbb C}^{K\times H_{f}}$ or a vector of
${\mathbb C}^{H_{f}}$) denote the learnable weight matrix and bias, respectively. Although the dimensionality of ${\bf B}^{({{\mathbb C}\rm FL})}_f$ may involve $K$, we can set the $K$ row vectors of ${\bf B}^{({{\mathbb C}\rm FL})}_f$ to be identical such that \eqref{cfl} is scalable to $K$. Furthermore, each ${\mathbb C}$FL is followed by a complex BatchNorm (BN) \cite{BatchNorm} layer to prevent overfitting and enhance the convergence learning behavior.

For the $F$-th ${\mathbb C}$FL, the update process is given by
\begin{flalign}
{\bf v}^{({{\mathbb C}\rm FL})}_{F} =  {\bf V}^{({{\mathbb C}\rm FL})}_{\left(F-1\right)} {\bf W}^{({{\mathbb C}\rm FL})}_{F} + {\bf b}^{({{\mathbb C}\rm FL})}_{F},
\end{flalign}
where ${\bf W}^{({{\mathbb C}\rm FL})}_{F} \in {\mathbb C}^{H_{(F-1)}\times H_F}$ and ${\bf b}^{({{\mathbb C}\rm FL})}_{F}\in {\mathbb C}^{1\times H_F}$ or $\in {\mathbb C}^{K\times H_F}$ denote the learnable parameters with $H_F$ being $N-1$, $1$, $1$ and $1$ for $F_1$, $F_2$, $F_3$ and $F_4$, respectively. Then, the outputs of the two stages are respectively obtained by
\begin{flalign}
&{\left[ {{\xi _2},{\xi _3},\ldots,{\xi _N}} \right]^T} = {\rm{Re}}\left( {{\bf{v}}_{{F_1}}^{({{\mathbb C}\rm{FL}})}} \right),\\
&{\xi _{\max }} = {\rm{Re}}\left( {{v}_{{F_2}}^{({{\mathbb C}\rm{FL}})}} \right),\\
&{\left[ {{p_1},{p_2},\ldots,{p_K}} \right]^T} =  P_{\max }\times{\rm Sigmod}\left({{\rm{Re}}\left( {{\bf{v}}_{{F_3}}^{({{\mathbb C}\rm{FL}})}} \right)}\right),\\
&{\left[ {{\alpha _1},{\alpha _2},\ldots,{\alpha _K}} \right]^T} = {\rm{Sigmod}}\left({\rm{Re}}\left( {{\bf{v}}_{{F_4}}^{({{\mathbb C}\rm{FL}})}} \right) \right).
\end{flalign}

At last, we can recover the desired solution $\left\{\mathbf{w}_i,\mathbf{x}\right\}$ to Problem \eqref{po2} following Section III-A.

Denote $\{{\mathbf{w}_i},{\mathbf{x}}| {\bm \Theta }\}$ by the output of the two-stage GNN with the input of $\{{{\theta}_i}\}$, where ${\bm \Theta }$ denotes all learnable parameters of the two-stage GNN
\begin{flalign}
{\bm \Theta}\triangleq&\left[{\bf a}^{({{\mathbb C}\rm
 GAL})}_{g,z}, {\bf W}^{({{\mathbb C}\rm
 GAL})}_{g,z}, {\widehat {\bf W}}^{({{\mathbb C}\rm
 GAL})}_{g,z}, {\overline {\bf a}}^{({{\mathbb C}\rm
 GAL})}_{g,z}, {\overline {\bf W}}^{({{\mathbb C}\rm
 GAL})}_{g,z}, \right. \nonumber  \\
 &{\widetilde {\bf W}}^{({{\mathbb C}\rm
 GAL})}_{g,z}, {\bf W}^{({{\mathbb C}\rm
 FL})}_{f}, {\bf W}^{({{\mathbb C}\rm
 FL})}_{F}, {\bf B}^{({{\mathbb C}\rm
 FL})}_{f}, {\bf b}^{({{\mathbb C}\rm
 FL})}_{F}].
 \end{flalign}
 It is observed that the dimensionality of $\bm\Theta$ is independent of the number of users, i.e., $K$. Therefore, by parameter sharing, the two-stage GNN is scalable to the number of users. That is, the two-stage GNN is acceptable to the unseen problem size during training phase.

\subsection{Unsupervised Loss Function}

 The two-stage GNN guarantees feasible solution. Therefore, we use the system utility function i.e., \eqref{suf1} or \eqref{suf2}, to construct the unsupervised loss function to update ${\bm \Theta}$, which is given by
  \begin{flalign}\label{Loss_Function}
 {{\cal L}_M}\left( {\bf{\Theta }} \right) = \frac{1}{M}\sum\nolimits_{m \in {\cal M}} {\frac{1}{{U{{\left( {\left\{ {{{\bf{w}}_i},{\bf{x}}|{\bf{\Theta }}} \right\}} \right)}^{\left( m \right)}}}}},
 \end{flalign}
 where $M$ denotes the batch size.

\section{Numerical Results}

{\emph{Simulation scenario.}}
The number of transmit antennas and the number of users are set as $(N,K)\in \{(4,2),(8,3),(8,4),(8,5)\}$. The power budget and the constant circuit power are set as $P_{\rm max}=1$ W and $P_{\rm C}=0.5$ W, respectively. The signal-to-noise ratio is set as $20$ dB. The central frequency is $1.8$ GHz, corresponding to a wavelength of $\lambda=0.167$ m. The antenna minimum spacing is set as $\Delta=\lambda/2$, and the adjustable area is of length $D=10\lambda$.

\emph{Dataset.} This work prepares $4$ datasets with different numbers of antennas and users as shown in Table \ref{Datasets}.

\begin{table}[t]
    \centering
    \caption{Datasets.}
    \label{Datasets}
    \begin{tabular}{c|c|c|c|c}
    \hline
    No. &$N$ &$K$ & Size& Type\\
     \hline
      \hline
     1   &4&2& 100,000& A\\
     \hline
     2   &8&3& 2,000& B\\
     \hline
     3   &8&4& 100,000& A\\
     \hline
     4   &8&5& 2,000& B\\
     \hline
    \end{tabular}
     \begin{tablenotes}
            \footnotesize
            \item {Type A: The training set, validation set and test set are all included.}
            \item {Type B: Only test set is included.}
    \end{tablenotes}
\end{table}

\emph{Implementation Details.} The learnable weights are initialized according to \cite{kaiming_normal} and the learning rate is initialized as $1 \times 10^{-6}$. The adaptive moment estimation  is adopted as the optimizer during the training phase. The batch size is set as $1,024$ for $2,000$ epochs of training with early stopping to prevent overfitting. Our implementation is developed using Python 3.11.10 with Pytorch 2.3.1 on a computer with Intel(R) Core(TM) i9-12900K CPU and NVIDIA RTX 3090 (24 GB of memory).

\emph{Baseline.}
In order to evaluate the pinching-antenna system and the proposed GNN  numerically, the following two baselines are considered, i.e.,
\begin{itemize}
\item {Equidistant-antenna system solved by convex optimization method}: A successive convex approximation based method,  termed ``CVX".
\item {FAS solved by MLP}:  A basic feed-forward  neural network, similar to \cite{FAS-MLP}, termed ``MLP".
\end{itemize}

\emph{Performance metrics.}
 The metrics to evaluate the DL models on the test sets are given as follows.
\begin{itemize}
  \item
  {Achievable sum rate/EE}: The average achievable sum rate/EE  by the DL model.
  \item
  {Scalability}: The average achievable sum rate/EE by the DL model with problem sizes unseen in the training phase.
  \item
  {Inference time}: The average running time required to calculate the feasible solution by the DL model.
\end{itemize}

\begin{table}[t]
\belowrulesep=0pt
\aboverulesep=0pt
\centering
\caption{Performance evaluation of sum rate [bit/s/Hz].}
\label{sr_performance}
\begin{tabular}{c|c|c||c|c|c}
 \hline
$N$ &$K_{\rm Tr}$&$K_{\rm Te}$ & CVX & MLP & Two-stage GNN \\
 \hline
 \hline
 \rowcolor{blue!10}
\cellcolor{white}4&\cellcolor{white}2&\cellcolor{white}2&{13.67}&{13.13}&\bf{14.92}\\
 \hline
  \multicolumn{3}{c||}{Inference time}& 1.23 s& \bf{0.010 ms}& 0.051 ms\\
 \hline
 \rowcolor{orange!10}
\cellcolor{white}~&\cellcolor{white}~&\cellcolor{white}3&{21.49}&{$\times$}&\bf{22.23}$^\dag$\\
 \rowcolor{blue!10}
\cellcolor{white}8&\cellcolor{white}4&\cellcolor{white}4&{25.49}&{14.33}&\bf{26.58}\\
\rowcolor{orange!10}
\cellcolor{white}~&\cellcolor{white}~&\cellcolor{white}5&{28.16}&{$\times$}&\bf{28.32}$^\dag$\\
 \hline
  \multicolumn{3}{c||}{Inference time}& 6.13 s& \bf{0.025 ms}& 0.059 ms\\
 \hline
\end{tabular}
\begin{tablenotes}
\footnotesize
\item $\times$ represents ``not applicable".
\item $K_{\rm {Tr}}$/$K_{\rm {Te}}$: Value of  $K$ in the training/test set.
\item{$^\dag$: The result marked with $^\dag$ represents the scalability performance, while the result without $^\dag$ represents the achievable sum rate.}
\end{tablenotes}
\end{table}

\begin{table}[t]
\belowrulesep=0pt
\aboverulesep=0pt
\centering
\caption{Performance evaluation of EE [bit/J/Hz].}
\label{ee_performance}
\begin{tabular}{c|c|c||c|c|c}
 \hline
$N$ &$K_{\rm Tr}$&$K_{\rm Te}$ & CVX & MLP & Two-stage GNN \\
 \hline
 \hline
 \rowcolor{blue!10}
\cellcolor{white}4&\cellcolor{white}2&\cellcolor{white}2&{13.29}&{7.03}&\bf{14.80}\\
 \hline
  \multicolumn{3}{c||}{Inference time}& 3.12 s& \bf{0.011 ms}& 0.051 ms\\
 \hline
 \rowcolor{orange!10}
\cellcolor{white}~&\cellcolor{white}~&\cellcolor{white}3&{21.98}&{$\times$}&\bf{22.15}$^\dag$\\
 \rowcolor{blue!10}
\cellcolor{white}8&\cellcolor{white}4&\cellcolor{white}4&{25.80}&{9.66}&\bf{26.17}\\
\rowcolor{orange!10}
\cellcolor{white}~&\cellcolor{white}~&\cellcolor{white}5&\bf{29.00}&{$\times$}&{26.69}$^\dag$\\
 \hline
  \multicolumn{3}{c||}{Inference time}& 5.92 s& \bf{0.026 ms}& 0.060 ms\\
  \hline
\end{tabular}
\begin{tablenotes}
\footnotesize
\item $\times$ represents ``not applicable".
\item{$^\dag$: The result marked with $^\dag$ represents the scalability performance, while the result without $^\dag$ represents the achievable EE.}
\end{tablenotes}
\end{table}

Tables \ref{sr_performance} and \ref{ee_performance} evaluate the proposed two-stage GNN in terms of the performance metrics. As seen, the two-stage GNN is able to achieves higher sum rate or EE than the baselines when facing seen problem sizes ($(N,K)=(4,2)$ and $(8,4)$) during the training phase. The performance gain is mainly due to the FAS, as the second stage is inferior slightly to the CVX in terms of optimality given the same antenna positions (cf. Table \ref{Ablation Study}). Moreover, the two-stage GNN is with much faster inference speed that the CVX. For the unseen problem sizes during the training phase, the two-stage GNN also works and outperforms the CVX in most cases. In a summary, the two-stage GNN enabled FAS realizes  real-time and scalable optimization with superior performance than traditional equidistant-antenna systems.

\begin{table}[t]
\belowrulesep=0pt
\aboverulesep=0pt
\centering
\caption{Effectiveness of two stages: $(N,K)=(8,4)$.}
\label{Ablation Study}
\begin{tabular}{c||c c|c c}
\hline
 \multirow{2}{*}{\diagbox [width=10em] {Beamforming}{Antenna}}  & \multicolumn{2}{c|}{Equidistant} & \multicolumn{2}{c}{First stage} \\
 \cline{2-5}
 & SR & EE & SR & EE \\
 \hline
 \hline
 CVX & 25.99 & 25.80 & 27.21 & 26.82 \\
 \hline
 Second stage & 25.03 & 24.78 & 26.58 & 26.17 \\
 \hline
\end{tabular}
\begin{tablenotes}
\footnotesize
\item SR stands for sum rate.
\end{tablenotes}
\end{table}

Table \ref{Ablation Study} validates the effectiveness of both stages and demonstrates that the two stages can separately implemented. For equidistant-antenna systems, the second stage (i.e., beamforming learning stage) is able to achieve close sum rate or EE to the CVX, but with much faster inference speed (cf. Tables \ref{sr_performance} and \ref{ee_performance}). For the FAS with antenna positions yielded by the first stage (i.e., antenna position stage), the sum rate or EE by both the CVX and the second stage is improved compared with the equidistant-antenna system.

\section{Conclusion}

This paper have proposed an unsupervised learning approach to optimize the FAS. We have introduced a two-stage GNN architecture to solve the system utility maximization problem with the first stage and the second stage yielding antenna position and beamforming vectors, respectively. The two-stage GNN has been shown to be scalable to number of users. Numerical results have demonstrated the advantage of the FAS in enhancing sum rate and EE, while the two-stage GNN was validated to be a promising approach for the FAS in terms of real-time and scalable optimization. Besides, each  stage can be independently implemented.

\end{document}